\documentclass[a4paper,11pt]{article}
\usepackage{amssymb}
\usepackage{amsmath,mathtools}
\usepackage{graphicx}
\usepackage{hyperref,enumitem}
\usepackage{multirow}
\usepackage[affil-it]{authblk}
\providecommand{\keywords}[1]
{
	\small    
	\textbf{\textit{Keywords--}} #1
}

\newcommand{\xdasharrow}[2][->]{
	\tikz[baseline=-\the\dimexpr\fontdimen22\textfont2\relax]{
		\node[anchor=south,font=\scriptsize, inner ysep=1.5pt,outer xsep=2.2pt](x){#2};
		\draw[shorten <=3.4pt,shorten >=3.4pt,dashed,#1](x.south west)--(x.south east);
	}
}
\newcommand{\bra}[1]{\ensuremath{\left\langle#1\right|}}
\newcommand{\ket}[1]{\ensuremath{\left|#1\right\rangle}}

\usepackage{tikz}
\usetikzlibrary{matrix,shapes,arrows,positioning,chains, calc}

\begin{document}
	
	\title{Measurement-Device-Independent Quantum Secure Direct Communication with User Authentication}
	
	\author{Nayana Das%
		\thanks{Email address: \texttt{dasnayana92@gmail.com} }}
	\affil{Applied Statistics Unit, Indian Statistical Institute, Kolkata, India.}
	
	\author{Goutam Paul %
		\thanks{Email address: \texttt{goutam.paul@isical.ac.in}}}
	\affil{Cryptology and Security Research Unit, R. C. Bose Centre for Cryptology and Security, Indian Statistical Institute, Kolkata, India.\\}
	
	\date{}
	
	\maketitle
	
	\maketitle
	
	\begin{abstract}
		Quantum secure direct communication (QSDC) and deterministic secure quantum communication (DSQC) are two important branches of quantum cryptography, where one can transmit a secret message securely without encrypting it by a prior key. In the practical scenario, an adversary can apply detector-side-channel attacks to get some non-negligible amount of information about the secret message. Measurement-device-independent (MDI) quantum protocols can remove this kind of detector-side-channel attacks, by introducing an untrusted third party (UTP), who performs all the measurements during the protocol with imperfect measurement devices. In this paper, we put forward the first MDI-QSDC protocol with user identity authentication, where both the sender and the receiver first check the authenticity of the other party and then exchange the secret message. Then we extend this to an MDI quantum dialogue (QD) protocol, where both the parties can send their respective secret messages after verifying the identity of the other party. Along with this, we also report the	first MDI-DSQC protocol with user identity authentication. Theoretical analyses prove the security of our proposed protocols against common attacks.
		
		\keywords{Collective attacks \and Deterministic secure quantum communication \and Identity authentication \and Measurement-device-independent \and Quantum cryptography \and Quantum dialogue}
		%\PACS{{03.67.Dd} {Quantum cryptography and communication security } \and {03.67.−a} {Quantum information}}
		% \subclass{MSC code1 \and MSC code2 \and more}
	\end{abstract}
	
	\section{Introduction}
	\label{intro}
	Quantum cryptography is an application of quantum mechanical properties into the field of cryptography, where the security does not depend on some mathematical hard problems. Here the fundamental principles of quantum mechanics are used to guarantee the unconditional communication security of the quantum cryptographic protocols. In 1984, Bennett and Brassard proposed the first quantum key distribution (QKD) protocol~\cite{bennett2020quantum}, based on Wiesner’s theory of quantum conjugate coding~\cite{wiesner1983conjugate}, and this is the first protocol of quantum cryptography. Since then, QKD has received extensive attention both theoretically~\cite{ekert1991quantum,bennett1992quantum1,long2002theoretically,li2016one} and experimentally~\cite{bennett1992experimental,zhao2006experimental,tang2014experimental,bedington2016nanosatellite,zhong2019proof}.
	
	\noindent{\bf  QSDC and DSQC:} 
	Besides QKD, quantum secure direct communication (QSDC)~\cite{long2002theoretically,deng2003two,deng2004secure,wang2005quantum,hu2016experimental,zhang2017quantum,xie2018semi,chen2018three,tao2019two,wu2019security,bebrov2020efficient,zhou2020device,Liu2020High,das2020cryptanalysis,das2021quantum,wang2021quantum,ye2021generic,long2021drastic,yin2021novel} and deterministic secure quantum communication (DSQC)~\cite{beige2002secure,qing2004deterministic,lucamarini2005secure,long2007quantum,xiu2009deterministic,hu2018deterministic,yuan2019deterministic,elsayed2020deterministic} are also two important primitives of quantum cryptography. {The basic difference between QKD and QSDC or DSQC is that, QKD is designed for generating the random keys between communication parties, while QSDC or DSQC is used for direct transmission of secret information}. Both QSDC and DSQC are used to securely transmit a secret message directly over a quantum channel, without using any pre-shared secret key for encryption and decryption. In QSDC, no other classical information is needed other than the security checking process, whereas, in DSQC, at least one bit of additional classical information is required to decode one qubit. 
	
	\noindent{\bf Quantum dialogue (QD):} It is a natural generalization of QSDC, i.e., a bidirectional QSDC, where both the parties can exchange their secret messages simultaneously through a quantum channel. In 2004 Nguyen proposed the first QD protocol~\cite{nguyen2004quantum} by generalizing the ping-pong-protocol~\cite{bostrom2002deterministic}. Over the past two decades, QD has gone through rapid developments~\cite{zhong2005quantum,xia2006quantum,xin2006secure,yan2007controlled,tan2008classical,gao2010two,maitra2017measurement,das2020two}. QSDC protocols for more than two parties are discussed in~\cite{gao2005deterministic,jin2006three,ting2005simultaneous,wang2006multiparty,gao2010cryptanalysis,tan2014multi,banerjee2018quantum,he2019multiparty,das2021secure}.
	
	\noindent{\bf MDI-QSDC:} However in practice, due to lack of perfect measurement devices, an adversary (Eve) can take advantage of this loophole of an imperfect measurement device and tries to steal information without being detected. In order to solve this problem, Lo et al. first proposed the concept of measurement-device-independent (MDI) QKD protocol~\cite{lo2012measurement}. In MDI protocols, a UTP performs all the measurements during the protocol using imperfect devices, and thus it removes all the detector side-channel attacks introduced by Eve~\cite{makarov2005faked,makarov2006effects,qi2005time,makarov2009controlling}. Using the same technique as MDI-QKD, Zhou et al. proposed the first MDI-QSDC protocol~\cite{zhou2020measurement}, and some other MDI-QSDC and MDI-QD protocols also proposed recently~\cite{niu2018measurement,wu2020high,niu2020security,zou2020measurement,gao2019long,das2020improving,maitra2017measurement,Liu2020High,das2020two}. Similar to MDI-QKD, in 2021 Yang et al. proposed the first MDI-DSQC protocol~\cite{yang2021high} based on the polarization-spatial-mode hyperencoded qudits.
	
	\noindent{\bf QSDC with authentication:} For any secure communication, identity authentication of each user is very necessary to defeat an impersonation attack. The first-ever quantum user identification scheme was proposed by Cr{\'e}peau et
	al.~\cite{crepeau1995quantum} in 1995. After that, Lee et al. proposed the first QSDC protocol with user authentication~\cite{lee2006quantum}. Later on, a number of new QSDC protocols with authentication are presented~\cite{zhang2007comment,dan2010new,chang2014controlled,hwang2014quantum,das2020cryptanalysis,das2021quantum}.
	
	\noindent{\bf Our contribution:} Here in this paper, we compose both the above concepts of MDI-QSDC and user identity authentication and present the first protocol of MDI-QSDC with user authentication. We extend our MDI-QSDC protocol to an MDI-QD protocol, which also provides user authentication. Then we also propose an MDI-DSQC protocol with user authentication and prove the security of all the above three protocols.
	
	\noindent{\bf Comparison with existing works:} We compare the efficiency of our proposed MDI-QSDC protocol with the existing works (see Table~\ref{Comparison}). { In~\cite{zhou2020measurement}, authors proposed an MDI-QSDC protocol based on the idea of quantum teleportation, where the sender prepares a Bell state and the receiver prepares a single qubit state. First, they do a Bell measurement, by UTP, to teleport the receiver's qubit to the sender, and then the sender encodes its secret message. To decode the secret message they do a single qubit measurement on $Z$ basis by UTP. Therefore the protocol~\cite{zhou2020measurement} requires three qubits and two measurements to communicate a single-bit message. In~\cite{niu2018measurement}, the authors proposed an MDI-QSDC protocol using entanglement swapping. To share a two-bit secret message, both the sender and the receiver prepare Bell states and perform entanglement swapping with the help of a third party. After that, the sender encodes the secret message. This protocol requires two Bell states and two Bell measurements for sending a two-bit message. In~\cite{das2020improving}, authors found a security loophole in~\cite{niu2018measurement} and proposed a modification over that. The modified version also requires the same resource as before. In~\cite{gao2019long}, the authors proposed a long-distance MDI-QSDC protocol by using ancillary entangled photon-pair sources and relay nodes. To transmit a single-bit message, they use two Bell states and a single qubit state. The protocol also requires two Bell measurements and a $Z$-basis measurement. Here in our present protocol, to send a two-bit message, we only use a Bell state and a Bell measurement. Therefore, on average it requires a qubit and half measurement to transfer a single-bit message. Also, none of the above existing works provide the user authentication feature before transferring the secret information.}
	
	\begin{table}[h]
		\centering
		\caption{Comparison between existing MDI-QSDC and our work}
		\resizebox{1\textwidth}{!}{
		\begin{tabular}{|c|c|c|c|c|}
			\hline
			\multirow{2}{*}{\textbf{Paper}} & \textbf{No. of qubits}   & \textbf{{No. of Bell Meas.}} & \textbf{{No. of S.B. Meas.}} & \textbf{User}           \\
			& \textbf{per message bit} & \textbf{{per message bit} }    &\textbf{{per message bit} }    & \textbf{authentication} \\ \hline
			Zhou et al.~\cite{zhou2020measurement}                 & $3$                      & {$1$}  &    {$1$}                    & No                      \\ \hline
			Neu et al.~\cite{niu2018measurement}                   & $2$                      & {$1$}  &     {$0$}                   & No                      \\ \hline
			Gao et al.~\cite{gao2019long}                        & $5$                      & {$2$}    &      {$1$}                  & No                      \\ \hline
			
			Das et al.~\cite{das2020improving}                   & $2$                      & {$1$}     &    {$0$}                   & No                      \\ \hline
			Present protocol                    & $1$ &     {${1}/{2}$ }                &  {$0$}                         & Yes                     \\ \hline
		\end{tabular}}
		\label{Comparison}
		\\
		\raggedright{~~{*Bell Meas.: Bell basis measurement, S.B. Meas.: Single basis measurement.}}
	\end{table}

	The rest of this paper is organized as follows: in Section~\ref{sec2}, we briefly describe our proposed MDI-QSDC with user authentication protocol and its security analysis. Then in the next section, we generalize MDI-QSDC protocol into an MDI-QD with user authentication protocol. Then Section~\ref{sec_mdi_dsqc} presents our MDI-DSQC protocol and finally Section~\ref{conclusion} concludes our results.

	\subsection*{Notations}
	Throughout the paper, we use some notations and we describe those common notations here.
	
	\begin{itemize}[label=$\bullet$]
		\item $Z$ basis $=\{\ket{0},\ket{1}\}$ basis.
		\item $\ket{+}=\frac{1}{\sqrt{2}}(\ket{0}+ \ket{1})$, $\ket{-}=\frac{1}{\sqrt{2}}(\ket{0}- \ket{1})$.
		\item $X$ basis $=\{\ket{+},\ket{-}\}$ basis.
		\item $I=\ket{0}\bra{0}+\ket{1}\bra{1}$.
		\item $\sigma_x=\ket{1}\bra{0}+\ket{0}\bra{1}$.
		\item $i\sigma_y=\ket{0}\bra{1}-\ket{1}\bra{0}$.
		\item $\sigma_z=\ket{0}\bra{0}-\ket{1}\bra{1}$.
		\item $H=\frac{1}{\sqrt{2}}(\sigma_x+\sigma_z)$ is the Hadamard operator.
		\item $\ket{\Phi^{+}}=\frac{1}{\sqrt{2}}(\ket{00}+ \ket{11})=\frac{1}{\sqrt{2}}(\ket{++}+ \ket{--})$. 
		\item $\ket{\Phi^{-}}=\frac{1}{\sqrt{2}}(\ket{00}- \ket{11})=\frac{1}{\sqrt{2}}(\ket{+-}+ \ket{-+})$.
		\item $\ket{\Psi^{+}}=\frac{1}{\sqrt{2}}(\ket{01}+ \ket{10})=\frac{1}{\sqrt{2}}(\ket{++}- \ket{--})$. 
		\item $\ket{\Psi^{-}}=\frac{1}{\sqrt{2}}(\ket{01}- \ket{10})=\frac{1}{\sqrt{2}}(\ket{+-}- \ket{-+})$.
		\item Bell basis $=\{\ket{\Phi^{+}},\ket{\Phi^{-}},\ket{\Psi^{+}},\ket{\Psi^{-}}\}$ basis.
		\item $S_i=i$-th element of finite sequence $S$.
		\item $S_{A,i}=i$-th element of finite sequence $S_A$.
		%\item $\bar{b}$ = bit complement of $b$.
		\item $\Pr(A)=$ Probability of occurrence of an event $A$.
		\item $\Pr(A|B)=$ Probability of occurrence of an event $A$ given that the event $B$ has already occurred.
	\end{itemize}
	
	\section{Proposed MDI-QSDC protocol with user authentication}\label{sec2}
	In this section, we propose our new MDI-QSDC protocol with user identity authentication process.
	
	Suppose Alice has an $n$-bit secret message $m$, which she wants to send Bob through a quantum channel with the help of some untrusted third-party (UTP), who performs all the measurements during the protocol. Alice and Bob have their secret user identities $Id_A$ and $Id_B$ (each of $2k$ bits) respectively, which they have shared previously by using some secured QKD. The protocol is as follows:
	
	\begin{enumerate}
		\item Alice chooses $c$ check bits and inserts those bits in random positions of $m$. Let the new bit string be $m'$ of length $n+c$. We assume this length to be even, i.e., $n+c=2N$ for some integer $N$.
		\item \label{state prep bob}\textbf{Bob:}
		\begin{enumerate}
			\item Prepares $(N+k)$ EPR pairs randomly in $\ket{\Phi^{+}}$, $\ket{\Phi^{-}}$, $\ket{\Psi^{+}}$ and $\ket{\Psi^{-}}$ states.
			He separates the entangled qubit pairs into two particle sequences $S_A$ and $S_B$ each of length $(N+k)$, where $S_A$ is formed by taking out one qubit from each pair, and the remaining partner qubits form $S_B$.
			
			\item He also prepares $k$ EPR pairs according to his identity $Id_B$. For $1 \leq i \leq k$, the $i$-th qubit pair $I_i$ is prepared as one of $\ket{\Phi^{+}}$,  $\ket{\Phi^{-}}$,  $\ket{\Psi^{+}}$ and $\ket{\Psi^{-}}$, if the value of $Id_{B,(2i-1)}Id_{B,2i}$ is one of $00$, $01$, $10$ and $11$ respectively. He creates two sequences $I_A$ and $I_B$ of single photons, such that for $1 \leq i \leq k$, the $i$-th qubits of $I_A$ and $I_B$ are partners of each other in the $i$-th EPR pair $I_i$.
			
			\item Bob chooses two sets $D_A$ and $D_B$, each of $d$ many decoy photons randomly prepared in $Z$-basis or $X$-basis. Then he randomly interleaves the qubits of $I_A (I_B)$ and $D_A (D_B)$ and $S_A (S_B)$ (maintaining the relative ordering of each set) to get a new sequence of single qubits $Q_A (Q_B)$ (i.e., $Q_P=S_P \cup I_P \cup D_P$, $P=A, B$).
			\item \label{Q_A Bob to Alice} Bob retains the $Q_B$-sequence and sends the $Q_A$-sequence to Alice through a quantum channel. 
			\item After Alice receives $Q_A$-sequence, Bob announces the positions of the qubits of $I_A$ and $D_A$.
		\end{enumerate}
		
		\item \textbf{Alice:}
		\begin{enumerate}
			\item She separates the qubits of $S_A$, $I_A$ and $D_A$ from $Q_A$. Then from the sequence $S_A$, she randomly chooses $N$ qubits to encode the secret message and the remaining $k$ qubits (say, the set $C_A$) are used to encode her secret identity $Id_A$. The encoding processes for $m'$ and $Id_A$ are the same. Alice encodes two bits of classical information into one qubit by applying an unitary operator. To encode $00, 01, 10$ and $11$, she applies the Pauli operators~\cite{nielsen2002quantum} $I$, $\sigma_x$, $i\sigma_y$ and $\sigma_{z}$ respectively. After encoding the classical information, let $S_A$ become $S_A'$.
			
			\item Alice randomly applies $I$, $\sigma_x$, $i\sigma_y$ and $\sigma_{z}$ on the qubits of $I_A$ and resulting in a new sequence $I_A'$. She randomly inserts the qubits of $I_A'$ into random positions of $S_A'$ and the new sequence be $Q_A'$.
			
			\item She randomly applies cover operations from $\{I,i\sigma_y,H,i\sigma_yH\}$ on the qubits of $D_A$, resulting in a new new sequence $D_A^1$.
			\item Alice sends $D_A^1$ sequence to UTP to check the security of the channel from Bob to Alice.
		\end{enumerate}
		\item After the UTP receives the sequence $D_A^1$, Bob announces the preparation bases of the qubits of $D_A$ and Alice announces the corresponding cover operations which she applies on those qubits.
		\item UTP measures the qubits of $D_A^1$ in proper bases and announces the measurement result. Note that if the cover operation belongs to the set $\{H,i\sigma_yH\}$, then UTP changes the basis to measure the corresponding qubit. For example, let the $i$-th qubit of $D_A$ be prepared in $Z$-basis and the $i$-th cover operation be $i\sigma_yH$, then UTP measures the $i$th qubit of $D_A^1$ in $X$-basis. From the measurement results, Alice and Bob calculate the error in the channel from Bob to Alice, and decide to continue or abort the protocol.
		
		\item Alice inserts a new set of $d'$ decoy photons $D_A'$ into random positions of $Q_A'$, resulting in a new sequence $Q_A''$. Alice sends $Q_A''$-sequence to UTP.
		
		\item Alice announces the positions and the preparation bases of the decoy qubits of $D_A'$. UTP measures the decoy qubits and publishes the measurement results, and from that Alice calculates the error in the quantum channel between Alice and UTP. If the estimated error is greater than some threshold value, then they terminate the protocol and otherwise go to the next step.
		
		\item Bob sends the sequence $Q_B$ to UTP and when all the qubits of $Q_B$ are reached to UTP, Bob announces the positions and the preparation bases of the decoy qubits of $D_B$. UTP measures those qubits in proper bases and discloses the measurement results, and Bob calculates the error in the quantum channel between Bob and UTP. If the estimated error is greater than some threshold value, then they terminate the protocol and otherwise go to the next step.
		
		\item \textbf{Authentication process:}
		\begin{enumerate}
			\item \label{authen-bob}Alice announces the positions of the qubits of $I_A'$ and Bob announces the positions of the qubits of $I_B$. For $1 \leq i \leq k$, UTP measures the $i$-th qubit pair $(I_{A,i}', I_{b,i})$ in Bell basis and announces the result. As Alice knows $Id_B$, she knows the exact state of each $I_i$, which is the joint state $I_{A,i}I_{B,i}$. Since she randomly applies Pauli operators on $I_{A,i}$, the joint state changes to $I_{A,i}'I_{B,i}$. Alice compares the measurement result with $I_{A,i}'I_{B,i}$ to confirm Bob's identity. If she finds a non-negligible error then she aborts the protocol.
			\item Alice announces the positions of the qubits of $C_A$  corresponding to her identity $Id_A$ and UTP measures those qubits with their partner qubits from $S_B$ (say, the set $C_B$) in Bell bases and announces the measurement result. Since Bob knows $Id_A$, he compares the measurement results with $Id_A$ and checks if Alice is a legitimate party or not.  If he finds a non-negligible error, he aborts the protocol.
		\end{enumerate}
		\item The UTP measures each qubit pair from $(S_A',S_B)$ in Bell basis and announces the measurement result. From the knowledge of $(S_A,S_B)$ and $(S_A',S_B)$, Bob decodes the classical bit string $m'$ {using Table~\eqref{MDI-QSDC table}}.
		\item Alice and Bob publicly compare the random check bits to check the integrity of the messages. If they find an acceptable error rate then Bob gets the secret message $m$ and the communication process is completed.	
	\end{enumerate}
	
	\begin{table}[h]
		\centering
		\renewcommand*{\arraystretch}{1.2}
		\caption{{Encoding and decoding rules of our proposed MDI-QSDC.}}
		\setlength{\tabcolsep}{10pt}
		\resizebox{1\textwidth}{!}{
			\begin{tabular}{|c|c|c|c|c|}
				\hline
				{ \textbf{Bob prepares}}            & { \textbf{Secret message}} & { \textbf{Alice's unitary}} & { \textbf{Final joint}}        & { \textbf{Decoded}}      \\
				{ \textbf{$(S_A,S_B)$}}             & { \textbf{bits of Alice}}  & { \textbf{$S_A$ to $S_A'$}} & { \textbf{state $(S_A',S_B)$}} & { \textbf{message bits}} \\ \hline
				{ }                                 & { $00$}                    & { $I$}                      & { $\ket{\Phi^+}$}              & { $00$}                  \\ \cline{2-5} 
				{ }                                 & { $01$}                    & { $\sigma_x$}               & { $\ket{\Psi^+}$}              & { $01$}                  \\ \cline{2-5} 
				{ }                                 & { $10$}                    & { $i\sigma_y$}              & { $\ket{\Psi^-}$}              & { $10$}                  \\ \cline{2-5} 
				\multirow{-4}{*}{{ $\ket{\Phi^+}$}} & { $11$}                    & { $\sigma_z$}               & { $\ket{\Phi^-}$}              & { $11$}                  \\ \hline
				{ }                                 & { $00$}                    & { $I$}                      & { $\ket{\Phi^-}$}              & { $00$}                  \\ \cline{2-5} 
				{ }                                 & { $01$}                    & { $\sigma_x$}               & { $\ket{\Psi^-}$}              & { $01$}                  \\ \cline{2-5} 
				{ }                                 & { $10$}                    & { $i\sigma_y$}              & { $\ket{\Psi^+}$}              & { $10$}                  \\ \cline{2-5} 
				\multirow{-4}{*}{{ $\ket{\Phi^-}$}} & { $11$}                    & { $\sigma_z$}               & { $\ket{\Phi^+}$}              & { $11$}                  \\ \hline
				{ }                                 & { $00$}                    & { $I$}                      & { $\ket{\Psi^+}$}              & { $00$}                  \\ \cline{2-5} 
				{ }                                 & { $01$}                    & { $\sigma_x$}               & { $\ket{\Phi^+}$}              & { $01$}                  \\ \cline{2-5} 
				{ }                                 & { $10$}                    & { $i\sigma_y$}              & { $\ket{\Phi^-}$}              & { $10$}                  \\ \cline{2-5} 
				\multirow{-4}{*}{{ $\ket{\Psi^+}$}} & { $11$}                    & { $\sigma_z$}               & { $\ket{\Psi^-}$}              & { $11$}                  \\ \hline
				{ }                                 & { $00$}                    & { $I$}                      & { $\ket{\Psi^-}$}              & { $00$}                  \\ \cline{2-5} 
				{ }                                 & { $01$}                    & { $\sigma_x$}               & { $\ket{\Phi^-}$}              & { $01$}                  \\ \cline{2-5} 
				{ }                                 & { $10$}                    & { $i\sigma_y$}              & { $\ket{\Phi^+}$}              & { $10$}                  \\ \cline{2-5} 
				\multirow{-4}{*}{{ $\ket{\Psi^-}$}} & { $11$}                    & { $\sigma_z$}               & { $\ket{\Psi^+}$}              & { $11$}                  \\ \hline
			\end{tabular}
		}\label{MDI-QSDC table}
	\end{table}
	
\begin{figure}[!htbp]
		\caption{Block diagram of the proposed MDI-QSDC with user authentication protocol}
		\resizebox{1\textwidth}{!}{
		\label{mdi-qsdc-protocol-block}
		\fbox{\includegraphics[scale=.4]{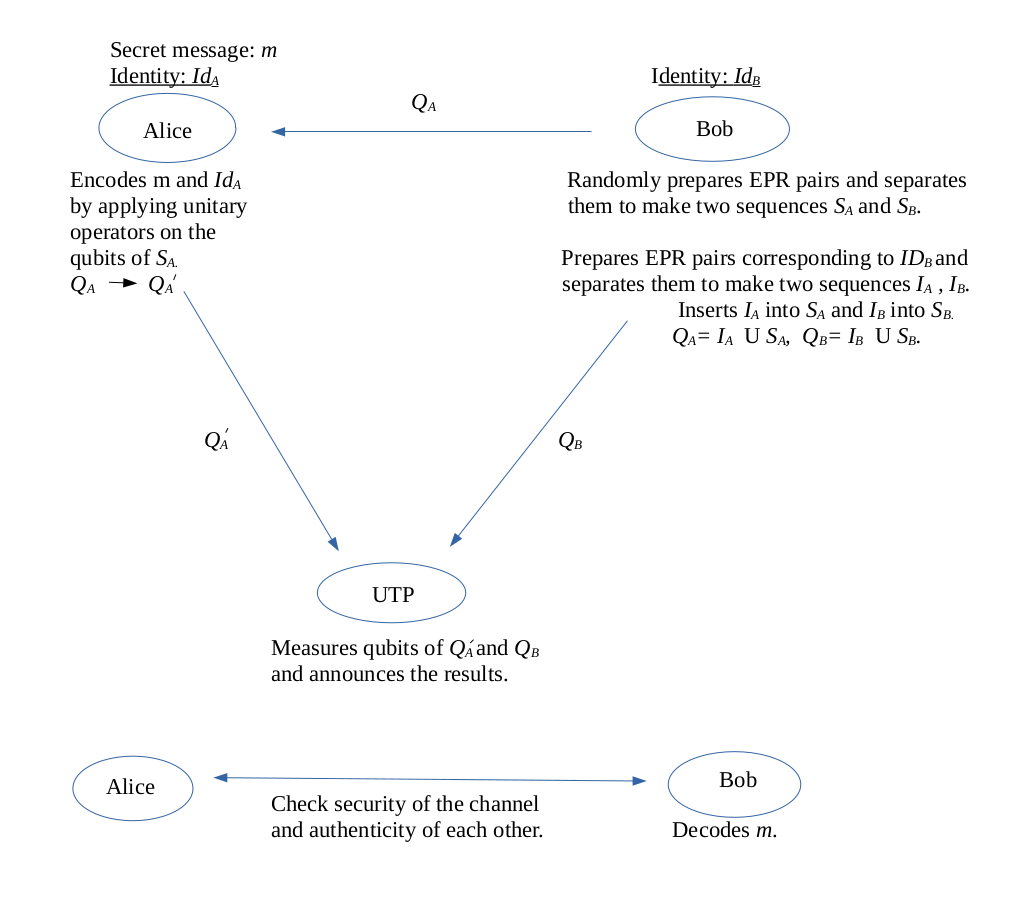}}}
	\end{figure}

	{Figure~\ref{mdi-qsdc-protocol-block} represents the block diagram of the proposed MDI-QSDC with user authentication protocol. We also present it in the form of an algorithm in figure~\ref{mdi-qsdc-protocol-table}, where we use the following notations.
		\begin{itemize}[label=$\bullet$]
			\item $X \rightarrow Y$: $X$ changes to $Y$.
			\item $\mathcal{P}(Q)$: Positions of the qubits of $Q$.
			\item $\mathcal{C}(Q)$: Cover operations on the qubits of $Q$.
			\item $\mathcal{B}(Q)$: Bases of the qubits of $Q$.
			\item $\mathcal{M}(Q) $ \& $\mathcal{A}$: Measures the qubits of $Q$ in proper bases and announces the results.
			\item $\mathcal{BM}(Q_1,Q_2)$ \& $\mathcal{A}$: Measures the qubit pairs of $(Q_1,Q_2)$ in Bell bases and announces the results.
			\item Sec.chk (A, B): Checks the security of the channel from A to B.
			\item Cov. op.: Cover operation. 
			\item Ins.: Inserts.
	\end{itemize}}
	
	\begin{figure}[!htbp]
		\caption{Proposed MDI-QSDC with user authentication protocol}
		\resizebox{1\textwidth}{!}{
			\centering
			\renewcommand*{\arraystretch}{2.4}
			\begin{tabular}{lllll}
				\hline
				
				\textbf{{\LARGE Alice}} (Identity $Id_A$)                                                     &                         & \textbf{{\LARGE UTP}}                      &                      & \textbf{{\LARGE Bob}}  (Identity $Id_B$)                                      \\
				\hline
				
				1. Ins. $c$ check bits into the                   &                         &                                   &                      &                                                    \\
				secret message $m$ and $m \rightarrow m'$.                                           &                         &                                   &                      &                                                    \\
				& \multicolumn{3}{l}{$\xdasharrow[<-,>=latex]{\text{~~~~~~~~~~~~~~~~~~					~~~~~~~~~~~~~~~~~~~~~~~~~~\textbf{2(d).}~$Q_A$~~~~~~~~~~~~~~~~~~~~~~~~~~~~~~~~~~~~~~~~~~~~~~~}}$} &      \textbf{2(a)-(c).}~Prepares $Q_A=S_A \cup I_A \cup D_A$       \\
				&                         &                                   &                      & and $Q_B=S_B \cup I_B \cup D_B$, where qubits          \\
				& \multicolumn{3}{l}{$\xleftarrow{\text{~~~~~~~~~~~~~~~~~~~~~~~~~~~~~~~~~~~~~~\textbf{2(e).}~$\mathcal{P}(I_A)$,~$\mathcal{P}(D_A)$~~~~~~~~~~~~~~~~~~~~~~~~~~~~~~~~~~~~~~~~~~~~~~~~}}$}                       & pair of $(S_A,S_B)$, $(I_A,I_B)$  are entangled,   \\
				&                         &          &                      & and qubits of $D_A$, $D_B$ are decoy states.       \\
				\textbf{3(a).}~Separates $S_A,I_A , D_A$ from $Q_A$.                           &                         &                                   &                      &                                                    \\
				Encodes $m',~Id_A$ on $S_A$, $S_A \rightarrow  S_A'$.                      &                         &                                   &                      &                                                    \\
				{$C_A$: qubits corresponding to $Id_A$.} & &           & & {$C_B$: partner qubits of $C_A$} \\
				\textbf{3(b).}~Random unitaries on $I_A$, $I_A \rightarrow I_A'$.                   &                         &                                   &                      &                                                    \\
				Ins. $I_A'$ into $S_A'$ and $Q_A'=S_A' \cup I_A'$.                 &                         &                                   &                      &                                                    \\
				\textbf{3(c).}~Cov. op. on $D_A$ and $D_A \rightarrow D_A^1$.                & $\xdasharrow{\text{~~~~~\textbf{3(d).}~$D_A^1$~~~~~~}}$                 &                                   &                      &                                                    \\
				& $\xrightarrow{\text{~~~~~~~\textbf{4$'$.}~$\mathcal{C}(D_A)$~~~~~~}}$    & \textbf{5.}~$\mathcal{M}(D_A^1) $ \& $\mathcal{A}$      & $\xleftarrow{\text{~~~~~~\textbf{4$'$.}~$\mathcal{B}(D_A)$~~~~~~~}}$         & \textbf{5$'$.}~Sec.chk (Bob, Alice)                \\
				\textbf{6.}~Ins. $D_A'$ into $Q_A'$, $Q_A' \rightarrow Q_A''$.                      &                         &                                   &                      &                                                    \\
				$D_A'$: set of new decoy states.                      & $\xdasharrow{\text{~~~~~~\textbf{6$'$.}~$Q_A''$ ~~~~~~}}$                 &                                   &                      &                                                    \\
				\textbf{7$''$.}~Sec.chk (Alice,UTP)                        & $\xrightarrow{\text{\textbf{7.}~$\mathcal{P}(D_A')$,~ $\mathcal{B}(D_A')$}}$   & \textbf{7$'$.}~$\mathcal{M}(D_A') $  \& $\mathcal{A}$      &                      &        \\
				&                         &                                   & $\xdasharrow[<-,>=latex]{\text{~~~~~~\textbf{8.}~$Q_B$~~~~~~~~~}}$              &                                                    \\
				&                         & \textbf{8$''$.}~$\mathcal{M}(D_B) $ \& $\mathcal{A}$ & $\xleftarrow{\text{\textbf{8$'$.}~$\mathcal{P}(D_B)$,~ $\mathcal{B}(D_B)$}}$  & \textbf{8$'''$.}~Sec.chk (Bob, UTP)  \\& & &     \\
				\textbf{9(a)$'''$.}~Verifies Bob's identity. & $\xrightarrow{\text{~~~\textbf{9(a).}~ $\mathcal{P}(I_A')$~~~}}$ & \textbf{9(a)$''$.}~$\mathcal{BM}(I_A',I_B)$ \& $\mathcal{A}$        & $\xleftarrow{\text{~~~\textbf{9(a)$'$.}~$\mathcal{P}(I_B)$, ~~~}}$      &                             \\
				&&&\\
				
				& $\xrightarrow{\text{~~\textbf{9(b).}~ $\mathcal{P}(C_A)$~~~~}}$ & \textbf{9(b)$'$.}~$\mathcal{BM}(C_A,C_B)$ \& $\mathcal{A}$   &    &  \textbf{9(b)$''$.}~Verifies Alice's identity.                              \\  & & &        \\
				&                         & \textbf{10.}~$\mathcal{BM}(S_A-C_A,S_B-C_B)$ \& $\mathcal{A}$          &                      & \textbf{10$'$.}~Decodes $m'$.                                   \\& & &    \\
				&    \multicolumn{3}{l}{$ \xleftrightarrow{\text{~~~~~~~~~~~~~~~~~~~~~~~~~~~~~~~~~~~~~\textbf{11.}~Compare the check bits~~~~~~~~~~~~~~~~~~~~~~~~~~~~~~~~~~~~~~~~~ }} $ }             & \textbf{11$'$.}~Extract $m$ from $m'$.  
				\\& & &   \\
				$\dashrightarrow$ denotes quantum channel,& & \\
				$\longrightarrow$ denotes classical channel.& &\\
				Step (i)$'$ happens just after Step (i).&&\\
				\hline                      
			\end{tabular}
		}
		\label{mdi-qsdc-protocol-table}
	\end{figure}

	\newpage
	\subsection{{Example of our MDI-QSDC protocol}}\label{example}
	
	{Let us now take an example of the above discussed MDI-QSDC with user authentication protocol, where we assume all channels are noiseless.}
	
	{Suppose Alice has a $6$-bit secret message $m=011010$ and the secret identities of Alice and Bob are $Id_A=1011$ and $Id_B=0111$ respectively, i.e., $n=6$ and $k=2$. Then the protocol is as follows.}
	
	\begin{enumerate}
		\item {Alice chooses $c=4$ check bits $1001$ and inserts those bits in random positions of $m$. Let the new bit string be $m'=0\mathbf{10}110\mathbf{0}1\mathbf{1}0$ (bold numbers are check bits, i.e., the 2nd, 3rd, 7th and 9th bits) of length $n+c=10=2N$, i.e., $N=5$.}
		
		\item \label{ex_state prep bob}\textbf{{Bob:}}
		\begin{enumerate}
			\item {Randomly prepares $N+k=7$ EPR pairs $$\ket{\Psi^{+}}_{a_1b_1}, \ket{\Phi^{+}}_{a_2b_2}, \ket{\Phi^{+}}_{a_3b_3}, \ket{\Psi^{-}}_{a_4b_4} \ket{\Phi^{-}}_{a_5b_5}, \ket{\Psi^{-}}_{a_6b_6}, \text{ and } \ket{\Psi^{+}}_{a_7b_7}.$$
				He separates the entangled qubit pairs into two particle sequences $$S_A=\{a_1,a_2,a_3,a_4,a_5,a_6,a_7\} \text{ and } S_B=\{b_1,b_2,b_3,b_4,b_5,b_6,b_7\},$$ each of length $7$.}
			
			\item {He also prepares $2$ EPR pairs $I_1=\ket{\Phi^{-}}_{a'_1b'_1} \text{ and } I_2=\ket{\Psi^{-}}_{'a_2b'_2}$ corresponding to his identity $Id_B=0111$, and creates two single-qubit sequences $I_A=\{a'_1,a'_2\}$ and $I_B=\{b'_1,b'_2\}$ by separating the EPR pairs.}
			
			\item {Bob chooses two sets $D_A=\{\ket{+},\ket{1},\ket{0},\ket{+}\}$ and $D_B=\{\ket{-},\ket{0},\ket{1},\ket{0}\}$, each of $d=4$ many decoy photons randomly prepared in $Z$-basis or $X$-basis. Then he randomly interleaves the qubits of $I_A (I_B)$ and $D_A (D_B)$ and $S_A (S_B)$ (maintaining the relative ordering of each set) to get a new sequences of single qubits $Q_A (Q_B)$. Let 
				$$Q_A=\{a_1,a_2,a'_1,\ket{+},a_3,\ket{1},a'_2,a_4,a_5,\ket{0},a_6,a_7,\ket{+}\}$$
				$$ \text{ and } Q_B=\{b_1,b'_1,b_2,b_3,b_4,\ket{-},\ket{0},b'_2,b_5,\ket{1},b_6,b_7,\ket{0}\}.$$}
			
			\item \label{ex_Q_A Bob to Alice} {Bob retains the $Q_B$-sequence and sends the $Q_A$-sequence to Alice through a quantum channel.} 
			
			\item {After Alice receives $Q_A$-sequence, Bob announces the positions of the qubits of $I_A$ (3rd and 7th) and $D_A$ (4th, 6th, 10th and 13th).}
		\end{enumerate}
		
		\item \textbf{{Alice:}}
		\begin{enumerate}
			\item {She separates the qubits of $S_A$, $I_A$ and $D_A$ from $Q_A$, i.e., she has $$S_A=\{a_1,a_2,a_3,a_4,a_5,a_6,a_7\}, I_A=\{a'_1,a'_2\} \text{ and } D_A=\{\ket{+},\ket{1},\ket{0},\ket{+}\}.$$
				She randomly chooses $5$ qubits $a_1,a_3,a_4,a_6$ and $a_7$ from $S_A$ to encode $m'=0101100110$ and the remaining $2$ qubits $a_2$ and $a_5$ (say, the set $C_A=\{a_2,a_5\}$) are used to encode $Id_A=1011$. After encoding the classical information, let $S_A$ become $S_A'$, then 
				$$S_A'=\{\sigma_x(a_1),i\sigma_y(a_2),\sigma_x(a_3),i\sigma_y(a_4),\sigma_{z}(a_5),\sigma_x(a_6),i\sigma_y(a_7)\}.$$}
			
			\item {Alice randomly applies $\sigma_z$ and $I$ on the qubits of $I_A$ and the resulting new sequence is  $I_A'=\{\sigma_{z}(a_1'),I(a_2')\}$. She randomly inserts the qubits of $I_A'$ into random positions of $S_A'$ and the new sequence is $$Q_A'=\{\sigma_x(a_1),\sigma_{z}(a_1'),i\sigma_y(a_2),\sigma_x(a_3),I(a_2'),i\sigma_y(a_4),\sigma_{z}(a_5),\sigma_x(a_6),i\sigma_y(a_7)\}.$$}
			
			\item {She randomly applies cover operations from $\{I,i\sigma_y,H,i\sigma_yH\}$ on the qubits of $D_A$ and the resulting new sequence is $$D_A^1=\{H(\ket{+}),i\sigma_yH(\ket{1}),i\sigma_y(\ket{0}),I(\ket{+})\}=\{\ket{0},\ket{+},\ket{1},\ket{+}\}.$$}
			
			\item {Alice sends $D_A^1$ to UTP to check the security of the channel from Bob to Alice.}
		\end{enumerate}
		
		\item {After the UTP receives the sequence $D_A^1$, Bob announces the preparation bases ($X,Z,Z$ and $X$) of the qubits of $D_A$ and Alice announces the corresponding cover operations ($H,i\sigma_yH ,i\sigma_y$ and $I$).}
		
		\item {UTP measures the qubits of $D_A^1$ in proper bases ($Z,X,Z$ and $X$) and announces the measurement results $\ket{0},\ket{+},\ket{1},\ket{+}$. Since there is no error, Alice and Bob continue the protocol.}
		
		\item {Alice prepares a new set of $d'=4$ decoy photons $D_A'=\{\ket{0},\ket{+},\ket{-},\ket{1}\}$. She inserts the decoy qubits into random positions of $Q_A'$ and sends the resulting new sequence $Q_A''$ to UTP, where
			$$Q_A''=\{\sigma_x(a_1), \sigma_{z}(a_1'), i\sigma_y(a_2), \ket{0}, \sigma_x(a_3), I(a_2'), \ket{+} , i\sigma_y(a_4), \ket{-},  \sigma_{z}(a_5) ,\sigma_x(a_6), \ket{1}, i\sigma_y(a_7)\}.$$ }
		
		\item {Alice announces the positions (4th, 7th, 9th and 12th) and the preparation bases ($Z,X,X$ and $Z$) of the decoy qubits of $D_A'$. UTP measures the decoy qubits and publishes the measurement results $\ket{0},\ket{+},\ket{-},\ket{1} $. Since there is no error, Alice and Bob continue the protocol.}
		
		\item {Bob sends the sequence $Q_B$ to UTP and when all the qubits of $Q_B$ are reached to UTP, Bob announces the positions (6th, 7th, 10th and 13th) and the preparation bases ($X,Z,Z$ and $Z$) of the decoy qubits of $D_B$. UTP measures those qubits in proper bases and discloses the measurement results $\ket{-},\ket{0},\ket{1},\ket{0}$. Then Bob calculates the error rate (which is zero for this example) in the quantum channel between Bob and UTP and goes to the next step.}
		
		\item \textbf{{Authentication process:}}
		\begin{enumerate}
			\item \label{ex_authen-bob}{Alice announces the positions (2nd and 6th) of the qubits of $I_A'$ in the sequence $Q_A''$ and Bob announces the positions (2nd and 8th) of the qubits of $I_B$ in the sequence $Q_B$. UTP measures the $i$-th qubit pairs $(\sigma_{z}(a_1'),b_1')$ and $(I(a_2'),b_2')$ in Bell basis and announces the results $\ket{\Phi^+}$ and $\ket{\Psi^-}$. As Alice knows $Id_B=0111$, she knows the exact states of $I_1=\ket{\Phi^-}$ and $I_2=\ket{\Psi^-}$. Since she randomly applied Pauli operators $\sigma_{z}, I$ on $a_1', a_2'$ respectively, the joint state changes to $\ket{\Phi^+}, \ket{\Psi^-}$. Alice confirms Bob's identity and continues the protocol.}
			
			\item {Alice announces the positions (2nd and 5th) of the qubits of $C_A$ in the sequence $S_A'$ and UTP measures those qubits with their partner qubits from $S_B$ (say, the set $C_B=(b_2,b_5)$) in Bell bases and announces the measurement results $\ket{\Psi^-}, \ket{\Phi^+}$. Since the initial states of the EPR pairs are $\ket{\Phi^+}, \ket{\Phi^-}$, Bob decodes the identity of Alice as $Id_A=1011$ and confirms Alice as a legitimate party and continues the protocol.}
		\end{enumerate}
		
		\item {The UTP measures each qubit pair from $(S_A',S_B)$ in Bell basis and announces the measurement result $\ket{\Phi^+}, \ket{\Psi^+}, \ket{\Phi^+},\ket{\Phi^-},\ket{\Phi^-}$. From these results, Bob decodes the classical bit string $m'=0101100110$.}
		
		\item {Alice and Bob publicly compare the random check bits (2nd, 3rd, 7th and 9th bits of $m'$) to check the integrity of the messages. Bob discards those bits to obtain the secret message $m=011010$ and the communication process is completed.}	
	\end{enumerate}

	\subsection{Security analysis of our MDI-QSDC protocol}\label{security}
	In our proposed MDI-QSDC with user authentication, the secret message is transmitted between two legitimate parties, and the
	potential adversary is kept ignorant of the content. There are also broadcast channels between Alice, Bob and UTP, for the necessary classical information, to execute the protocol. First, we show the security of our proposed MDI-QSDC protocol for user authentication by establishing the security against impersonation attack. Then we prove the security of the message transmission part.
	
	\subsubsection{Security for user authentication}
	Let us now discuss the security of our proposed MDI-QSDC protocol against impersonation attacks. An eavesdropper, Eve, may try to impersonate Alice in order to send a fake message to Bob. But since Eve does not know the pre-shared key $Id_A$, Bob can easily detect Eve with a very high probability. In the proposed MDI-QSDC protocol, suppose Eve may intercept the sequence $Q_A$ sent from Bob to Alice in Step~\ref{Q_A Bob to Alice}. However, without knowing the
	pre-shared key $Id_A$, Eve applies Pauli operators randomly on $k$ qubits of $C_A$, instead of performing the correct unitary to encode $Id_A$. She sends it to UTP, who measures these qubits with their partner qubits from $C_B$ on the Bell basis and announces the results. Since Bob knows the initial state of those $k$ EPR pairs $(C_A,C_B)$ and the value of $Id_A$, he compares the measurement results with the expected EPR pairs and detects Eve. Since Eve applies Pauli operators randomly on each qubit, she applies correct unitary with probability $\frac{1}{4}$ and hence the detection probability of Bob is $1-(\frac{1}{4})^k$. 
	
	On the other hand, Eve may try to impersonate Bob to get the secret message from Alice. In the proposed MDI-QSDC protocol, suppose Eve initiates the protocol and generates the sequences of qubits $Q_A$ and $Q_B$, which contain the sequences $I_A$ and $I_B$ respectively, by following the process described in Step~\ref{state prep bob}. Now, since Eve does not know the value of $Id_B$, she prepares each $I_i$ ($1 \leq i \leq k$) as one of the EPR pairs randomly with probability $\frac{1}{4}$. After Alice applies cover operations on the qubits of $I_A$, the set becomes $I_A'$. In the authentication process (Step~\ref{authen-bob}), UTP measures the joint states of $(I_A',I_B)$ in proper bases and announces the results. As Alice knows the value of $Id_B$, she compares the measurement results with the expected results and detects Eve with probability $1-(\frac{1}{4})^k$.

	\subsubsection{Security for message transmission}
	In our MDI-QSDC protocol, we are ignorant of the measurement process
	and strategy that an adversary may exploit, hence we focus on the system after Bob sends the sequence $Q_A$ to Alice, where a joint state 
	$\rho^{jnt}_{AB}$, consisting of maximally entangled photon pairs shared between Alice and Bob. We consider a situation where an adversary Eve attacks the system with an auxiliary system and performs a coherent attack. Here, in our protocol, Alice and Bob use decoy states to obtain the gain and quantum bit error rate (QBER) after each transmission of qubits sequences where both of them send single qubits to the UTP. Now we use the concept of virtual qubits~\cite{gottesman2004security,lo2012measurement} and the proof technique of~\cite{niu2020security} to establish the security of our protocol against this type of attack. The idea of virtual qubit is that, instead of preparing a single qubit decoy state from $\{\ket{0},\ket{1},\ket{+}\,\ket{-}\}$, Alice (Bob) prepares EPR pair, which is a combined system of her (his) virtual qubit and the qubit she (he) is sending to the UTP. Subsequently, they measure their virtual qubits to decide to continue or abort the protocol. For simplicity, let us assume that initially Bob prepares all the EPR pairs in $\ket{\Phi^+}$ and he applies the cover operations $I, \sigma_z,\sigma_x,i\sigma_y$ on the qubits of $S_B$ while sending this sequence $Q_B$ to the UTP. Note that this step is equivalent to the fact that Bob prepares EPR pairs randomly from the set of all Bell states.
	
	Let the system of Alice, Bob and Eve be $A$, $B$ and $E$ respectively. Then from Csiszár–Körner theory~\cite{csiszar1978broadcast}, the secrecy capacity between Alice and Bob is $C_S$, 
	\begin{equation}
	C_S=\max [I(A:B)-I(A:E)],
	\end{equation}
	where $I(X:Y)$ stands for mutual information of two random variables $X$ and $Y$. Now if $C_S > 0$, then there is a forward encoding scheme with a capacity less than $C_S$, which can be used to transmit the message reliably and securely from Alice to Bob.
	
	According to quantum De Finetti representation theorem~\cite{renner2007symmetry}, the joint state $\rho^{jnt}_{AB}$ can be asymptotically approximated as a direct product of independent and identically distributed (i.i.d.) subsystems $\rho^{\otimes N}_{AB}$, if a randomized permutation is applied to the system. Thus Eve attacks each qubit separately by using a separate probe $\ket{E}$ {and then the coherent attack model can be considered as the collective attack by Eve}. 
	
	According to~\cite{kraus2005lower},  $\rho_{AB}$ can be written as a linear combination of the Bell states as follows, 
	\begin{equation}
	\rho_{AB}=\delta_1\ket{\Phi^+}\bra{\Phi^+}+ \delta_2\ket{\Phi^-}\bra{\Phi^-}+ \delta_3\ket{\Psi^+}\bra{\Psi^+}+ \delta_4\ket{\Psi^-}\bra{\Psi^-},
	\end{equation} 
	where $\sum^4_{i=1}\delta_i=1$.
	Let $\ket{\Phi_{ABE}}$ be a purification of the mixed state $\rho_{AB}$. Then it can be written as
	\begin{equation}
	\ket{\Phi_{ABE}}=\sum^4_{i=1}\sqrt{\delta_i}\ket{\Psi_i}\ket{E_i},
	\end{equation}
	where $\ket{\Psi_1}=\ket{\Phi^+}$, $\ket{\Psi_2}=\ket{\Phi^-}$, $\ket{\Psi_3}=\ket{\Psi^+}$, $\ket{\Psi_4}=\ket{\Psi^-}$ are the entangled pairs shared by Alice and Bob, and $\ket{E_i}$, $1 \leq i \leq 4$, are the orthonormal states of the system $\ket{E}$. 
	
	After Bob sends the sequence $Q_A$ to Alice, they calculate the bit error rate $\epsilon_z$ and phase error rate $\epsilon_x$ by measuring the virtual qubits by Bob and their partner qubits by Alice. They choose the same bases, either $(Z,Z)$ or $(X,X)$ with probability $\frac{1}{2}$, and measure their respective qubits. If no error occurs, then they should get the same outcomes as $\ket{\Phi^{+}}=\frac{1}{\sqrt{2}}(\ket{00}+ \ket{11})=\frac{1}{\sqrt{2}}(\ket{++}+ \ket{--})$. If they get different outcomes while measuring in $Z$-basis, i.e., the shared entangled state is either $\ket{\Psi^+}$ or $\ket{\Psi^-}$, then bit flip error occurs and thus $\epsilon_z=\delta_3+\delta_4$. Similarly, when they measure in $X$-basis and get different outcomes, phase error occurs and thus $\epsilon_x=\delta_2+\delta_4$.	 
	If both the error rates are less than some predefined threshold value, then they continue the process and Alice encodes her message by applying proper unitary operators $U_{\zeta}$'s on the qubits of $S_A$ and Bob applies random cover operations from the set of all Pauli operators on the qubits of $S_B$, and send their respective sequences to the UTP. Then the shared state becomes
	\begin{equation}
	\begin{split}
	\rho_{ABE}^{\zeta} & = \frac{1}{4}U_{\zeta}(\ket{\Phi_{ABE}}\bra{\Phi_{ABE}}+ \sigma_z^{B}\ket{\Phi_{ABE}}\bra{\Phi_{ABE}}\sigma_z^{B} \\
	& +\sigma_x^{B}\ket{\Phi_{ABE}}\bra{\Phi_{ABE}}\sigma_x^{B}-\sigma_y^{B}\ket{\Phi_{ABE}}\bra{\Phi_{ABE}}\sigma_y^{B})U_{\zeta}^\dagger\\
	&=U_{\zeta}\rho_{ABE}^{c}U_{\zeta}^\dagger,
	\end{split}
	\end{equation}
	where $\zeta \in \{00,01,10,11\}$ and $U_{00}=I$, $U_{01}=\sigma_x$, $U_{10}=i\sigma_y$, $U_{11}=\sigma_z$ are the message encoding operations of Alice, and $\rho_{ABE}^{c}= \frac{1}{4}(\ket{\Phi_{ABE}}\bra{\Phi_{ABE}}+ \sigma_z^{B}\ket{\Phi_{ABE}}\bra{\Phi_{ABE}}\sigma_z^{B}+\sigma_x^{B}\ket{\Phi_{ABE}}\bra{\Phi_{ABE}}\sigma_x^{B}-\sigma_y^{B}\ket{\Phi_{ABE}}\bra{\Phi_{ABE}}\sigma_y^{B}).$
	
	%	\begin{equation}
	%	\begin{split}
	%	\rho_{ABE}^{c} & = \frac{1}{4}(\ket{\Phi_{ABE}}\bra{\Phi_{ABE}}+ \sigma_z^{B}\ket{\Phi_{ABE}}\bra{\Phi_{ABE}}\sigma_z^{B} \\
	%	& +\sigma_x^{B}\ket{\Phi_{ABE}}\bra{\Phi_{ABE}}\sigma_x^{B}-\sigma_y^{B}\ket{\Phi_{ABE}}\bra{\Phi_{ABE}}\sigma_y^{B}).
	%	\end{split}
	%	\end{equation}
	%	Let the $2N$-bit message of Alice be $m'=\zeta_1\zeta_2\ldots \zeta_N$, where for $1 \leq i \leq N$, $\zeta_i$ is a two-bit binary number randomly chosen from $\{00,01,10,11\}$. Then the probability distribution of each $\zeta_i$ is $p_{\zeta_i}=\frac{1}{4}$. For $1 \leq i \leq N$, Alice encodes $\zeta_i$ by applying $U_{\zeta_i}$ on $\rho_{ABE}^{c}$ and the state becomes $\rho_{ABE}^{\zeta_i}$. We now calculate the maximum amount of accessible information of Eve about $\zeta_i$ for the subsystem $AE$. Then from Holevo theorem~\cite{holevo1973bounds} we can see the mutual information 
	%	$I(A:E)$ is bounded above as follows,
	
	Let the $2N$-bit message of Alice be $m'=\zeta_1\zeta_2\ldots \zeta_N$, where for $1 \leq i \leq N$, $\zeta_i$ is a two-bit binary number randomly chosen from $\mathcal{B}=\{00,01,10,11\}$ and the probability distribution of each $\zeta_i$ is $\frac{1}{4}$. For $1 \leq i \leq N$, Alice encodes $\zeta_i$ by applying $U_{\zeta_i}$ on $\rho_{ABE}^{c}$ and the state becomes $\rho_{ABE}^{\zeta_i}$. We now calculate the maximum amount of accessible information of Eve about $\zeta_i$. Then from Holevo theorem~\cite{holevo1973bounds}, we see the mutual information $I(A:E)$ is bounded above as,
	
	\begin{equation}\label{eq_mutual_info}
	\begin{split}
	I(A:E)& \leq S\left( \sum_{\zeta \in \mathcal{B}}p_{\zeta} \rho_{ABE}^{\zeta}\right) - \sum_{\zeta \in \mathcal{B}}p_{\zeta}S(\rho_{ABE}^{\zeta})\\
	%& \leq S\left( \sum_{\zeta}p_{\zeta} \rho_{ABE}^{\zeta}\right)-2,
	\end{split}
	\end{equation}
	where $p_{\zeta}=\frac{1}{4}$, the probability of randomly selecting one element from $\mathcal{B}$, and $S(\cdot)$ is the Von Neumann entropy.
	
	One can see that Alice's encoding and Bob's cover operations make a maximal mixture of the subsystems $A$ and $B$. 	Thus we have $S(\rho_{ABE}^{\zeta})=2$ for $\zeta \in \mathcal{B}$, and 
	\begin{equation}\label{eq_mutual_info1}
	I(A:E) \leq S\left( \sum_{\zeta}p_{\zeta} \rho_{ABE}^{\zeta}\right)-2,
	\end{equation} 
	and
	\begin{equation}\label{channel}
	\begin{split}
	\sum_{\zeta}p_{\zeta} \rho_{ABE}^{\zeta} &=\rho^{mix}_{AB}\otimes Tr_{AB}(\ket{\Phi_{ABE}}\bra{\Phi_{ABE}})\\
	&=\rho^{mix}_{AB}\otimes \sum_{j=1}^4 \delta_j\ket{E_j}\bra{E_j},
	\end{split}
	\end{equation}
	where $\rho^{mix}_{AB}=\frac{I}{4}$ is the maximally mixed state of the system $AB$.
	Now we have from Equation~\eqref{channel},
	\begin{equation}\label{eq8}
	\begin{split}
	S\left( \sum_{\zeta}p_{\zeta} \rho_{ABE}^{\zeta}\right)& =S\left(\rho^{mix}_{AB}\otimes \sum_{j=1}^4\delta_j\ket{E_j}\bra{E_j} \right) \\ 
	&=S(\rho^{mix}_{AB}) +S\left( \sum_{j=1}^4 \delta_j \ket{E_j}\bra{E_j}\right) \\
	&=S\left(\frac{I}{4} \right) +  \sum_{j=1}^4 \delta_j\log{\frac{1}{\delta_j}}\\
	&=2 + H(\delta_j),
	\end{split}
	\end{equation}
	where $H(\cdot)$ represents the Shannon entropy function.
	
	{\textbf{Lemma 1:} For a probability distribution $\{\delta_i, 1\leq i \leq 4\}$, $-\sum_{i=1}^4 \delta_i log \delta_i  \leq h(\delta_2+ \delta_4)+ h(\delta_3+ \delta_4)$, where $h(\cdot)$ represents the binary entropy function. (See appendix for proof.)}
	
	{Then from Equation~\eqref{eq_mutual_info1} and Equation~\eqref{eq8},
		\begin{equation}
		\begin{split}
		I(A:E) \leq H(\delta_j)& = \sum_{j=1}^4 \delta_j\log{\frac{1}{\delta_j}}\\
		&\leq h(\delta_3+\delta_4)+h(\delta_2+\delta_4) \text{ (by Lemma~1)}\\
		&=h(\epsilon_z)+h(\epsilon_x),
		\end{split}
		\end{equation}}

	Let $\epsilon_e$ be the error rate calculated after message decoding step, and if there is a discrete symmetric channel between Alice and Bob, then the secrecy capacity is 
	\begin{equation*}
	\begin{split}
	C_S &\geq I(A:B)-I(A:E)\\
	&\geq H(A)-H(A|B)- h(\epsilon_z)-h(\epsilon_x)\\
	&= 2-h(\epsilon_e) -h(\epsilon_z)-h(\epsilon_x).
	\end{split}
	\end{equation*} 
	For our protocol to be secure, we need $C_S > 0$, i.e., $2-h(\epsilon_e)>h(\epsilon_z)+h(\epsilon_x)$.
	
	In the next two sections, we propose MDI-QD and MDI-DSQC protocols with mutual identity authentication respectively.
	
	\section{Proposed MDI-QD protocol with user authentication}\label{sec3}
	In this section, we generalize the MDI-QSDC protocol into an MDI-QD protocol, which also provides mutual user authentication. Here, both Alice and Bob send their $n$-bit secret message to each other simultaneously after confirming the authenticity of the other user. They use one EPR pair to exchange one-bit messages from each other. Bob randomly prepares $(n+c)$ EPR pairs $\ket{\Phi^+}$ or $\ket{\Psi^+}$ ($\ket{\Phi^-}$ or $\ket{\Psi^-}$) corresponding to his secret message bit $0$ ($1$), where $c$ is the number of check bits. {He also randomly prepares $k$ EPR pairs from $\{\ket{\Phi^+},\ket{\Phi^-},\ket{\Psi^+},\ket{\Psi^-}\}$ for encoding the secret identity of Alice and inserts these into the previously prepared EPR sequence. After Alice receives the qubit sequence, he announces the positions of randomly prepared EPR pairs}. Alice randomly applies Pauli operator $I$ or $\sigma_{z}$ ($\sigma_x$ or $i\sigma_y$) to encode her message bit $0$ ($1$) (see Table~\eqref{MDI-QD table}). The rest of the procedure is the same as the above MDI-QSDC protocol described in Section~\ref{sec2}. The security of this protocol directly follows from the above MDI-QSDC protocol. 
	
	\begin{table}[h]
		\centering
		\renewcommand*{\arraystretch}{1.2}
		\caption{Encoding rules of our proposed MDI-QD.}
		\setlength{\tabcolsep}{10pt}
		\resizebox{1\textwidth}{!}{
			\begin{tabular}{|c|c|c|c|c|}
				\hline
				\multicolumn{2}{|c|}{\textbf{Message bit}}  & \textbf{Bob prepares}           & \textbf{Alice's unitary} & \textbf{Final joint state} \\ \hline
				\textbf{Alice}       & \textbf{Bob}         & \textbf{$(S_A,S_B)$}            & \textbf{$S_A$ to $S_A'$} & \textbf{$(S_A',S_B)$}      \\ \hline
				\multirow{4}{*}{$0$} & \multirow{4}{*}{$0$} & \multirow{2}{*}{$\ket{\Phi^+}$} & $I$                      & $\ket{\Phi^+}$             \\ \cline{4-5} 
				&                      &                                 & $\sigma_z$               & $\ket{\Phi^-}$             \\ \cline{3-5} 
				&                      & \multirow{2}{*}{$\ket{\Psi^+}$} & $I$                      & $\ket{\Psi^+}$             \\ \cline{4-5} 
				&                      &                                 & $\sigma_z$               & $\ket{\Psi^-}$             \\ \hline
				\multirow{4}{*}{$0$} & \multirow{4}{*}{$1$} & \multirow{2}{*}{$\ket{\Phi^-}$} & $I$                      & $\ket{\Phi^-}$             \\ \cline{4-5} 
				&                      &                                 & $\sigma_z$               & $\ket{\Phi^+}$             \\ \cline{3-5} 
				&                      & \multirow{2}{*}{$\ket{\Psi^-}$} & $I$                      & $\ket{\Psi^-}$             \\ \cline{4-5} 
				&                      &                                 & $\sigma_z$               & $\ket{\Psi^+}$             \\ \hline
				\multirow{4}{*}{$1$} & \multirow{4}{*}{$0$} & \multirow{2}{*}{$\ket{\Phi^+}$} & $\sigma_x$               & $\ket{\Psi^+}$             \\ \cline{4-5} 
				&                      &                                 & $i\sigma_y$              & $\ket{\Psi^-}$             \\ \cline{3-5} 
				&                      & \multirow{2}{*}{$\ket{\Psi^+}$} & $\sigma_x$               & $\ket{\Phi^+}$             \\ \cline{4-5} 
				&                      &                                 & $i\sigma_y$              & $\ket{\Phi^-}$             \\ \hline
				\multirow{4}{*}{$1$} & \multirow{4}{*}{$1$} & \multirow{2}{*}{$\ket{\Phi^-}$} & $\sigma_x$               & $\ket{\Psi^-}$             \\ \cline{4-5} 
				&                      &                                 & $i\sigma_y$              & $\ket{\Psi^+}$             \\ \cline{3-5} 
				&                      & \multirow{2}{*}{$\ket{\Psi^-}$} & $\sigma_x$               & $\ket{\Phi^-}$             \\ \cline{4-5} 
				&                      &                                 & $i\sigma_y$              & $\ket{\Phi^+}$             \\ \hline
			\end{tabular}
		}\label{MDI-QD table}
	\end{table}
	
	\subsection{{Example of our MDI-QD protocol}}\label{example_qd}
	
	{Let us now take an example of the above discussed MDI-QD with user authentication protocol, where we assume all channels are noiseless.}
	
	{Suppose Alice (Bob) has the $3$-bit secret message $m_a=011$ ($m_b=100$) and $4$-bit secret identity $Id_A=1011$ ($Id_B=0111$), i.e., $n=3$ and $k=2$. Then the protocol is as follows.}
	
	\begin{enumerate}
		\item {Alice (Bob) chooses $c=2$ check bits $10$ ($01$) and inserts those bits in random positions of $m_a$ ($m_b$). Let the new bit string be $m_a'=\mathbf{1}01\mathbf{0}1$ ($m_b'=1\mathbf{0}0\mathbf{1}0$) of length $5$, where the bold numbers represent the check bits.}
		
		\item \label{ex_state prep bob_qd}\textbf{{Bob:}}
		\begin{enumerate}
			\item {Prepares $5$ EPR pairs corresponding to $m_b'$ and those are $$\ket{\Psi^{-}}_{a_1b_1},  \ket{\Phi^{+}}_{a_3b_3}, \ket{\Psi^{+}}_{a_4b_4} , \ket{\Phi^{-}}_{a_6b_6}, \text{ and } \ket{\Phi^{+}}_{a_7b_7}.$$ 
				He separates the entangled qubit pairs into two particle sequences $$S_A=\{a_1,a_3,a_4,a_6,a_7\} \text{ and } S_B=\{b_1,b_3,b_4,b_6,b_7\},$$ each of length $5$.}
			
			\item {He also randomly prepares $2$ EPR pairs 	$\ket{\Phi^{+}}_{a_2b_2}$ and $\ket{\Phi^{-}}_{a_5b_5}$ and separates into two particle sequences $C_A=\{a_2,a_5\}$ and $C_B=\{b_2,b_5\}$. He inserts the qubits of $C_A$ and $C_B$ into the sequences $S_A$ and $S_B$ to form two new sequences $$S_A'=\{a_1,a_2,a_3,a_4,a_5,a_6,a_7\} \text{ and } S_B'=\{b_1,b_2,b_3,b_4,b_5,b_6,b_7\}$$  respectively.}
			
			\item {Then he prepares $2$ EPR pairs $I_1=\ket{\Phi^{-}}_{a'_1b'_1} \text{ and } I_2=\ket{\Psi^{-}}_{'a_2b'_2}$ corresponding to his identity $Id_B=0111$, and creates two single-qubit sequences $I_A=\{a'_1,a'_2\}$ and $I_B=\{b'_1,b'_2\}$ by separating the EPR pairs.}
			
			\item {Bob chooses two sets $D_A=\{\ket{+},\ket{1},\ket{0},\ket{+}\}$ and $D_B=\{\ket{-},\ket{0},\ket{1},\ket{0}\}$, each of $d=4$ many decoy photons randomly prepared in $Z$-basis or $X$-basis. Then he randomly interleaves the qubits of $I_A (I_B)$ and $D_A (D_B)$ and $S_A' (S_B')$ (maintaining the relative ordering of each set) to get a new sequences of single qubits $Q_A (Q_B)$. Let 
				$$Q_A=\{a_1,a_2,a'_1,\ket{+},a_3,\ket{1},a'_2,a_4,a_5,\ket{0},a_6,a_7,\ket{+}\}$$
				$$ \text{ and } Q_B=\{b_1,b'_1,b_2,b_3,b_4,\ket{-},\ket{0},b'_2,b_5,\ket{1},b_6,b_7,\ket{0}\}.$$}
			
			\item \label{ex_Q_A Bob to Alice_qd} {Bob retains the $Q_B$-sequence and sends the $Q_A$-sequence to Alice through a quantum channel.} 
			
			\item {After Alice receives $Q_A$-sequence, Bob announces the positions of the qubits of $C_A$ (2nd and 9th), $I_A$ (3rd and 7th) and $D_A$ (4th, 6th, 10th and 13th).}
		\end{enumerate}
		
		\item \textbf{{Alice:}}
		\begin{enumerate}
			\item {She separates the qubits of $S_A$, $C_A$, $I_A$ and $D_A$ from $Q_A$, i.e., she has $$S_A=\{a_1,a_3,a_4,a_6,a_7\}, C_A=\{a_2,a_5\}, I_A=\{a'_1,a'_2\} \text{ and } D_A=\{\ket{+},\ket{1},\ket{0},\ket{+}\}.$$
				She encodes $m_a'=10101$ and $Id_A=1011$ on the qubits of $S_A$ and $C_A$ respectively. After encoding the classical information, let $S_A$ and $C_A$ become $S_A^1$ and $C_A^1$ respectively. Then 
				$$S_A^1=\{\sigma_x(a_1),\sigma_z(a_3),i\sigma_y(a_4),I(a_6),i\sigma_y(a_7)\}$$ and 
				$$C_A^1=\{i\sigma_y(a_2),\sigma_{z}(a_5)\}.$$
				Then she randomly inserts the qubits of $C_A^1$ into the $S_A^1$ and let the new sequence be 
				$$S_A''=\{\sigma_x(a_1),i\sigma_y(a_2),\sigma_z(a_3),i\sigma_y(a_4),\sigma_{z}(a_5),I(a_6),i\sigma_y(a_7)\}.$$
			}
			
			\item {Alice randomly applies $\sigma_z$ and $I$ on the qubits of $I_A$ and the resulting new sequence is  $I_A'=\{\sigma_{z}(a_1'),I(a_2')\}$. She randomly inserts the qubits of $I_A'$ into random positions of $S_A''$ and the new sequence is $$Q_A'=\{\sigma_x(a_1),\sigma_{z}(a_1'),i\sigma_y(a_2),\sigma_z(a_3),I(a_2'),i\sigma_y(a_4),\sigma_{z}(a_5),I(a_6),i\sigma_y(a_7)\}.$$}
			
			\item {She randomly applies cover operations from $\{I,i\sigma_y,H,i\sigma_yH\}$ on the qubits of $D_A$ and the resulting new sequence is $$D_A^1=\{H(\ket{+}),i\sigma_yH(\ket{1}),i\sigma_y(\ket{0}),I(\ket{+})\}=\{\ket{0},\ket{+},\ket{1},\ket{+}\}.$$}
			
			\item {Alice sends $D_A^1$ to UTP to check the security of the channel from Bob to Alice.}
		\end{enumerate}
		
		\item {After the UTP receives the sequence $D_A^1$, Bob announces the preparation bases ($X,Z,Z$ and $X$) of the qubits of $D_A$ and Alice announces the corresponding cover operations ($H,i\sigma_yH ,i\sigma_y$ and $I$).}
		
		\item {UTP measures the qubits of $D_A^1$ in proper bases ($Z,X,Z$ and $X$) and announces the measurement results $\ket{0},\ket{+},\ket{1},\ket{+}$. Since there is no error, Alice and Bob continue the protocol.}
		
		\item {Alice prepares a new set of $d'=4$ decoy photons $D_A'=\{\ket{0},\ket{+},\ket{-},\ket{1}\}$. She inserts the decoy qubits into random positions of $Q_A'$ and sends the resulting new sequence $Q_A''$ to UTP, where
			$$Q_A''=\{\sigma_x(a_1), \sigma_{z}(a_1'), i\sigma_y(a_2), \ket{0}, \sigma_z(a_3), I(a_2'), \ket{+} , i\sigma_y(a_4), \ket{-},  \sigma_{z}(a_5) ,I(a_6), \ket{1}, i\sigma_y(a_7)\}.$$ }
		
		\item {Alice announces the positions (4th, 7th, 9th and 12th) and the preparation bases ($Z,X,X$ and $Z$) of the decoy qubits of $D_A'$. UTP measures the decoy qubits and publishes the measurement results $\ket{0},\ket{+},\ket{-},\ket{1} $. Since there is no error, Alice and Bob continue the protocol.}
		
		\item {Bob sends the sequence $Q_B$ to UTP and when all the qubits of $Q_B$ are reached to UTP, Bob announces the positions (6th, 7th, 10th and 13th) and the preparation bases ($X,Z,Z$ and $Z$) of the decoy qubits of $D_B$. UTP measures those qubits in proper bases and discloses the measurement results $\ket{-},\ket{0},\ket{1},\ket{0}$. Then Bob calculates the error rate (which is zero for this example) in the quantum channel between Bob and UTP and goes to the next step.}
		
		\item \textbf{{Authentication process:}}
		\begin{enumerate}
			\item \label{ex_authen-bob_qd}{Alice announces the positions (2nd and 6th) of the qubits of $I_A'$ in the sequence $Q_A''$ and Bob announces the positions (2nd and 8th) of the qubits of $I_B$ in the sequence $Q_B$. UTP measures the $i$-th qubit pairs $(\sigma_{z}(a_1'),b_1')$ and $(I(a_2'),b_2')$ in Bell basis and announces the results $\ket{\Phi^+}$ and $\ket{\Psi^-}$. As Alice knows $Id_B=0111$, she knows the exact states of $I_1=\ket{\Phi^-}$ and $I_2=\ket{\Psi^-}$. Since she randomly applied Pauli operators $\sigma_{z}, I$ on $a_1', a_2'$ respectively, the joint state changes to $\ket{\Phi^+}, \ket{\Psi^-}$. Alice confirms Bob's identity and continues the protocol.}
			
			\item {Alice announces the positions (2nd and 5th) of the qubits of $C_A'$ in the sequence $S_A''$ and UTP measures those qubits with their partner qubits from $C_B=(b_2,b_5)$ in Bell bases and announces the measurement results $\ket{\Psi^-}, \ket{\Phi^+}$. Since the initial states of the EPR pairs are $\ket{\Phi^+}, \ket{\Phi^-}$, Bob decodes the identity of Alice as $Id_A=1011$ and confirms Alice as a legitimate party and continues the protocol.}
		\end{enumerate}
		
		\item {The UTP measures each qubit pair from $(S_A',S_B)$ in Bell basis and announces the measurement result $\ket{\Phi^-}, \ket{\Phi^-},\ket{\Phi^-},\ket{\Phi^-}, \ket{\Psi^-}$. From these results, Alice (Bob) decodes the classical bit string $m_b'=10010$ ($m_a'=10101$).}
		
		\item {Alice and Bob publicly compare the random check bits to check the integrity of the messages. They discard those bits to obtain the secret message $m_a=011$ and $m_b=100$. This completes the communication process.}	
	\end{enumerate}

	\section{Proposed MDI-DSQC Protocol with user authentication}\label{sec_mdi_dsqc}
	In this section, we propose our new MDI-DSQC protocol with user identity authentication process.
	
	Let Alice has an $n$-bit secret message $m$, which she wants to send Bob through a quantum channel with the help of some UTP, who performs all the measurements during the protocol. Alice and Bob have their $2k$-bit secret user identities $Id_A$ and $Id_B$ respectively which they have shared previously by using some secured QKD. The protocol is as follows:
	
	Steps 1, 2, 3(a) are the same as before in the MDI-DSQC protocol of Section~\ref{sec2}.
	\begin{enumerate}
		\setcounter{enumi}{2}
		
		\item \textbf{Alice:} 
		\begin{enumerate}
			\item She separates the qubits of $S_A$, $I_A$ and $D_A$ from $Q_A$. Then from the sequence $S_A$ she randomly chooses $N$ qubits to encode the secret message and the remaining $k$ qubits are used to encode her secret identity $Id_A$. The encoding processes for $m'$ and $Id_A$ are the same. Alice encodes two bits of classical information into one qubit by applying an unitary operator. To encode $00, 01, 10$ and $11$ she applies the Pauli operators~\cite{nielsen2002quantum} $I$, $\sigma_x$, $i\sigma_y$ and $\sigma_{z}$ respectively. After encoding the classical information, suppose $S_A$ becomes $S_A'$.
			\item Alice randomly applies $I$, $\sigma_x$, $i\sigma_y$ and $\sigma_{z}$ on the qubits of $I_A$ to get, say, $I_A'$. She randomly inserts the qubits of $I_A'$ and $D_A$ into random positions of $S_A'$ and let the new sequence be $Q_A'$.
			\item She randomly applies cover operations from $\{I,i\sigma_y,H,i\sigma_yH\}$ on the qubits of $Q_A'$ and inserts a new set of $d'$ decoy photons $D_A'$ into random positions of $Q_A'$, to obtain, say, $Q_A''$, which Alice sends to UTP.
		\end{enumerate}
		\item After UTP receives the sequence $Q_A''$, Alice announces the positions and the preparation bases of the decoy qubits of $D_A'$. UTP measures the decoy qubits and publishes the measurement results, and Alice calculates the error in the quantum channel between Alice and UTP. If the estimated error is greater than some threshold value, then they terminate the protocol and otherwise go to the next step.
		\item Bob sends the sequence $Q_B$ to UTP and when all the qubits of $Q_B$ are reached to UTP, Bob announces the positions and the preparation bases of the decoy qubits of $D_B$. UTP measures those qubits in proper bases and discloses the measurement results, and Bob calculates the error in the quantum channel between Bob and UTP. If the estimated error is greater than some threshold value, then they terminate the protocol and otherwise go to the next step.
		\item To check the security of the quantum channel from Bob to Alice, Bob announces the preparation bases of the qubits of $D_A$ and Alice announces the corresponding positions and the cover operations which she applies on those qubits. UTP measures those qubits, from the announced measurement results Alice and Bob calculate the error in the channel and decide to continue or stop the protocol.
		\item UTP discards all the measured qubits and Alice announces all cover operations for the remaining qubits.
		\item \textbf{Authentication process:}
		Same as before in the MDI-DSQC protocol of Section~\ref{sec2}.
		\item UTP measures each qubit pair from $(S_A',S_B)$ in Bell basis and announces the measurement result. From the knowledge of $(S_A,S_B)$ and $(S_A',S_B)$, Bob decodes the classical bit string $m'$.
		\item Alice and Bob publicly compare the random check bits to check the integrity of the messages. If they find an acceptable error rate then Bob gets the secret message $m$ and the communication process is completed.	
	\end{enumerate}
	
	Using similar arguments as in Section~\ref{security}, we can prove the security of our proposed MDI-DSQC Protocol with user authentication.
	
	\subsection{{Example of our MDI-DSQC protocol}}\label{example_dsqc}
	
	{Let us now take an example of the above discussed MDI-DSQC with user authentication protocol, where we assume all channels are noiseless.}
	
	{Suppose Alice has a $6$-bit secret message $m=011010$ and the secret identities of Alice and Bob are $Id_A=1011$ and $Id_B=0111$ respectively, i.e., $n=6$ and $k=2$. Then the protocol is as follows.}
	
	\begin{enumerate}
		\item {Alice chooses $c=4$ check bits $1001$ and inserts those bits in random positions of $m$. Let the new bit string be $m'=0\mathbf{10}110\mathbf{0}1\mathbf{1}0$ (bold numbers are check bits, i.e., the 2nd, 3rd, 7th and 9th bits) of length $n+c=10=2N$, i.e., $N=5$.}
		
		\item \label{ex_dsqc_state prep bob}\textbf{{Bob:}}
		\begin{enumerate}
			\item {Randomly prepares $N+k=7$ EPR pairs $$\ket{\Psi^{+}}_{a_1b_1}, \ket{\Phi^{+}}_{a_2b_2}, \ket{\Phi^{+}}_{a_3b_3}, \ket{\Psi^{-}}_{a_4b_4} \ket{\Phi^{-}}_{a_5b_5}, \ket{\Psi^{-}}_{a_6b_6}, \text{ and } \ket{\Psi^{+}}_{a_7b_7}.$$
				He separates the entangled qubit pairs into two particle sequences $$S_A=\{a_1,a_2,a_3,a_4,a_5,a_6,a_7\} \text{ and } S_B=\{b_1,b_2,b_3,b_4,b_5,b_6,b_7\},$$ each of length $7$.}
			
			\item {He also prepares $2$ EPR pairs $I_1=\ket{\Phi^{-}}_{a'_1b'_1} \text{ and } I_2=\ket{\Psi^{-}}_{'a_2b'_2}$ corresponding to his identity $Id_B=0111$, and creates two single-qubit sequences $I_A=\{a'_1,a'_2\}$ and $I_B=\{b'_1,b'_2\}$ by separating the EPR pairs.}
			
			\item {Bob chooses two sets $D_A=\{\ket{+},\ket{1},\ket{0},\ket{+}\}$ and $D_B=\{\ket{-},\ket{0},\ket{1},\ket{0}\}$, each of $d=4$ many decoy photons randomly prepared in $Z$-basis or $X$-basis. Then he randomly interleaves the qubits of $I_A (I_B)$ and $D_A (D_B)$ and $S_A (S_B)$ (maintaining the relative ordering of each set) to get a new sequences of single qubits $Q_A (Q_B)$. Let 
				$$Q_A=\{a_1,a_2,a'_1,\ket{+},a_3,\ket{1},a'_2,a_4,a_5,\ket{0},a_6,a_7,\ket{+}\}$$
				$$ \text{ and } Q_B=\{b_1,b'_1,b_2,b_3,b_4,\ket{-},\ket{0},b'_2,b_5,\ket{1},b_6,b_7,\ket{0}\}.$$}
			
			\item \label{ex_dsqc_Q_A Bob to Alice} {Bob retains the $Q_B$-sequence and sends the $Q_A$-sequence to Alice through a quantum channel.} 
			
			\item {After Alice receives $Q_A$-sequence, Bob announces the positions of the qubits of $I_A$ (3rd and 7th) and $D_A$ (4th, 6th, 10th and 13th).}
		\end{enumerate}
		
		\item \textbf{{Alice:}}
		\begin{enumerate}
			\item {She separates the qubits of $S_A$, $I_A$ and $D_A$ from $Q_A$, i.e., she has $$S_A=\{a_1,a_2,a_3,a_4,a_5,a_6,a_7\}, I_A=\{a'_1,a'_2\} \text{ and } D_A=\{\ket{+},\ket{1},\ket{0},\ket{+}\}.$$
				She randomly chooses $5$ qubits $a_1,a_3,a_4,a_6$ and $a_7$ from $S_A$ to encode $m'=0101100110$ and the remaining $2$ qubits $a_2$ and $a_5$ (say, the set $C_A=\{a_2,a_5\}$) are used to encode $Id_A=1011$. After encoding the classical information, let $S_A$ become $S_A'$, then 
				$$S_A'=\{\sigma_x(a_1),i\sigma_y(a_2),\sigma_x(a_3),i\sigma_y(a_4),\sigma_{z}(a_5),\sigma_x(a_6),i\sigma_y(a_7)\}.$$}
			
			\item {Alice randomly applies $\sigma_z$ and $I$ on the qubits of $I_A$ and the resulting new sequence is  $I_A'=\{\sigma_{z}(a_1'),I(a_2')\}$. She randomly inserts the qubits of $I_A'$ and $D_A$ into random positions of $S_A'$ and the new sequence is $$Q_A'=\{\sigma_x(a_1),\ket{+},\sigma_{z}(a_1'),i\sigma_y(a_2),\ket{1},\ket{0},\sigma_x(a_3),I(a_2'),i\sigma_y(a_4),\ket{+},\sigma_{z}(a_5),\sigma_x(a_6),i\sigma_y(a_7)\}.$$}
			
			\item {She randomly applies cover operations from $\{I,i\sigma_y,H,i\sigma_yH\}$ on the qubits of $Q_A'$ and the resulting new sequence is
				\begin{gather*}
				{Q_A'}^1=\{i\sigma_yH\sigma_x(a_1),H(\ket{+}),I\sigma_{z}(a_1'),Hi\sigma_y(a_2),I(\ket{1}),i\sigma_y(\ket{0}),H\sigma_x(a_3), \\
				HI(a_2'),i\sigma_yHi\sigma_y(a_4),I(\ket{+}),i\sigma_y\sigma_{z}(a_5),i\sigma_yH\sigma_x(a_6),Hi\sigma_y(a_7)\}.
				\end{gather*}
				Alice choses a set $D_A'=\{\ket{-},\ket{1},\ket{0}\}$ of $d'=3$ decoy qubits randomly prepared in $Z$-basis or $X$-basis. Then she inserts those decoy qubits into some random positions of $Q_A'$ and the resulting new sequence is 
				\begin{gather*}
				Q_A''=\{\ket{-},i\sigma_yH\sigma_x(a_1),H(\ket{+}),I\sigma_{z}(a_1'),Hi\sigma_y(a_2),I(\ket{1}),\ket{1},i\sigma_y(\ket{0}),H\sigma_x(a_3), \\
				HI(a_2'),i\sigma_yHi\sigma_y(a_4),I(\ket{+}),i\sigma_y\sigma_{z}(a_5),i\sigma_yH\sigma_x(a_6),\ket{0},Hi\sigma_y(a_7)\}.
				\end{gather*}
				Alice sends $Q_A''$ to UTP.}
		\end{enumerate}
		
		\item {After the UTP receives the sequence $Q_A''$, Alice announces the positions (1st, 7th and 15th) and the preparation bases ($X,Z$ and $Z$) of the decoy qubits of $D_A'$. UTP measures the decoy qubits and publishes the measurement	results $\ket{-},\ket{1},\ket{0}$. Since there is no error, the quantum channel between Alice and UTP is secure and they continue the protocol.}
		
		\item {Bob sends the sequence $Q_B$ to UTP and when all the qubits of $Q_B$ are reached to UTP, Bob announces the positions (6th, 7th, 10th and 13th) and the preparation bases ($X,Z,Z$ and $Z$) of the decoy qubits of $D_B$. UTP measures those qubits in proper bases and discloses the measurement results $\ket{-},\ket{0},\ket{1},\ket{0}$. Then Bob calculates the error rate (which is zero for this example) in the quantum channel between Bob and UTP and goes to the next step.}
		
		\item {Bob announces the preparation bases ($X,Z,Z$ and $X$) of the qubits of $D_A$ and Alice announces the corresponding positions (3rd, 6th, 8th and 12th) in the sequence ${Q_A''}$ and the cover operations ($H, I, i\sigma_y$ and $I$) which she applies on those qubits. UTP measures those qubits and from the announced measurement results, Alice and Bob find the channel is secure. They decide to continue the protocol.}
		
		\item {UTP discards all the measured qubits from $Q_A''$ and $Q_B$, then UTP has the following sequences
			\begin{gather*}
			Q_A^1=\{i\sigma_yH\sigma_x(a_1),I\sigma_{z}(a_1'),Hi\sigma_y(a_2),H\sigma_x(a_3),HI(a_2'), i\sigma_yHi\sigma_y(a_4),\\ i\sigma_y\sigma_{z}(a_5), i\sigma_yH\sigma_x(a_6), Hi\sigma_y(a_7)\}
			\end{gather*} 
			and 
			\begin{gather*}
			Q_B^1=\{b_1,b'_1,b_2,b_3,b_4,b'_2,b_5,b_6,b_7\}.
			\end{gather*}
			Alice announces all cover operations ($i\sigma_yH, I, H, H, H, i\sigma_yH, i\sigma_y, i\sigma_yH$ and $H$) for the qubits of $Q_A^1$. Then UTP applies the inverse of the cover operation on the corresponding qubits and gets back 
			\begin{gather*}
			Q_A^2=\{\sigma_x(a_1),\sigma_{z}(a_1'),i\sigma_y(a_2),\sigma_x(a_3),I(a_2'), i\sigma_y(a_4),\sigma_{z}(a_5), \sigma_x(a_6), i\sigma_y(a_7)\}.
			\end{gather*} }
		
		\item \textbf{{Authentication process:}}
		\begin{enumerate}
			\item \label{ex_DSQC_authen-bob}{Alice announces the positions (2nd and 5th) of the qubits of $I_A'$ in the sequence $Q_A^2$ and Bob announces the positions (2nd and 6th) of the qubits of $I_B$ in the sequence $Q_B^1$. UTP measures the qubit pairs $(\sigma_{z}(a_1'),b_1')$ and $(I(a_2'),b_2')$ in Bell basis and announces the results $\ket{\Phi^+}$ and $\ket{\Psi^-}$. As Alice knows $Id_B=0111$, she knows the exact states of $I_1=\ket{\Phi^-}$ and $I_2=\ket{\Psi^-}$. Since she randomly applied Pauli operators $\sigma_{z}, I$ on $a_1', a_2'$ respectively, the joint state changes to $\ket{\Phi^+}, \ket{\Psi^-}$. Alice confirms Bob's identity and continues the protocol.}
			
			\item {Alice announces the positions (2nd and 5th) of the qubits of $C_A$ in the sequence $S_A'$ and UTP measures those qubits with their partner qubits from $S_B$ (say, the set $C_B=(b_2,b_5)$) in Bell bases and announces the measurement results $\ket{\Psi^-}, \ket{\Phi^+}$. Since the initial states of the EPR pairs are $\ket{\Phi^+}, \ket{\Phi^-}$, Bob decodes the identity of Alice as $Id_A=1011$ and confirms Alice as a legitimate party and continues the protocol.}
		\end{enumerate}
		
		\item {The UTP discards the measured qubits and measures the remaining qubit pairs from $(S_A',S_B)$ in Bell basis and announces the measurement result $\ket{\Phi^+}, \ket{\Psi^+}, \ket{\Phi^+},\ket{\Phi^-},\ket{\Phi^-}$. From these results, Bob decodes the classical bit string $m'=0101100110$.}
		
		\item {Alice and Bob publicly compare the random check bits (2nd, 3rd, 7th and 9th bits of $m'$) to check the integrity of the messages. Bob discards those bits to obtain the secret message $m=011010$ and the communication process is completed.}	
	\end{enumerate}

	\section{Conclusion}\label{conclusion}
	In this paper, we report the first-ever protocol for MDI-QSDC which provides mutual identity authentication of the users. Here, both the parties have their previously shared secret identity keys, and the sender first verifies the authenticity of the receiver and then sends the secret message with the help of a UTP, who performs all the measurements. Similarly, the receiver also verifies the sender's identity before receiving the message. Then we extend it to an MDI-QD protocol, where both the parties check the authenticity of the other party before exchanging their secret messages. Next, we also present an MDI-DSQC protocol with user authentication and analyses the security of these protocols.

	{	
		
		\section*{Appendix: Proof of Lemma~1}
		\textbf{Lemma 1:} For a probability distribution $\{\delta_i, 1\leq i \leq 4\}$, $-\sum_{i=1}^4 \delta_i log \delta_i  \leq h(\delta_2+ \delta_4)+ h(\delta_3+ \delta_4)$, where $h(\cdot)$ represents the binary entropy function.\vspace{.5cm}
		\\ 
		\textbf{Proof:} Let $X$ be a random variable such that
		\begin{equation*}
		{X} =
		\begin{cases}
		00 &\text{with probability $\delta_1$},\\
		01 &\text{with probability $\delta_2$},\\
		10 &\text{with probability $\delta_3$},\\
		11 &\text{with probability $\delta_4$}.
		\end{cases}
		\end{equation*}
		
		Let $Y$ and $Z$ be the following events,
		\begin{equation*}
		{Y} =
		\begin{cases}
		1, &\text{if the least significant bit of $X=1$ },\\
		0, &\text{otherwise}.
		\end{cases}
		\end{equation*}
		
		\begin{equation*}
		{Z} =
		\begin{cases}
		1, &\text{if the most significant bit of $X=1$ },\\
		0, &\text{otherwise}.
		\end{cases}
		\end{equation*}
		
		In other words,
		\begin{equation*}
		{Y} =
		\begin{cases}
		1 &\text{with probability $\delta_2+ \delta_4$ },\\
		0 &\text{with probability $\delta_1+ \delta_3$}.
		\end{cases}
		\end{equation*}
		and
		\begin{equation}
		{Z} =
		\begin{cases}
		1 &\text{with probability $\delta_3 + \delta_4$ },\\
		0 &\text{with probability $\delta_1+ \delta_2$}.
		\end{cases}
		\end{equation}
		
		Then the entropy of the events $Y$ and $Z$ are as follows 
		$$H(Y)=-\sum_{y \in \{0,1\}} \Pr(Y=y)log[\Pr(Y=y)]=h(\delta_2+ \delta_4).$$
		
		$$H(Z)=-\sum_{z \in \{0,1\}} \Pr(Z=z)log[\Pr(Z=z)]=h(\delta_3+ \delta_4).$$
		
		The joint entropy $H(Y,Z)$ of the events $Y$ and $Z$ is 
		\begin{equation*}
		\begin{split}
		H(Y,Z)& = -\sum_{y \in \{0,1\}} \sum_{z \in \{0,1\}} \Pr(Y=y,Z=z)log[\Pr(Y=y,Z=z)] \\
		&=-\sum_{x \in \{00,01,10,11\}} \Pr(X=x)log[\Pr(X=x)]\\
		&=-\sum_{i=1}^4 \delta_i log \delta_i.
		\end{split}
		\end{equation*}
		
		Now using sub-additivity property of entropy, i.e., the fact that the joint entropy of a set of variables is less than or equal to the sum of the individual entropies of the variables in the set. Therefore,
		
		\begin{equation*}
		\begin{split}
		H(Y,Z)& \leq  H(Y)+H(Z) \\
		\text{or, } -\sum_{i=1}^4 \delta_i log \delta_i & \leq h(\delta_2+ \delta_4)+ h(\delta_3+ \delta_4).
		\end{split}
		\end{equation*}
	}
	\bibliographystyle{unsrt}
	\bibliography{main}
	
\end{document}